# A Peer-to-Peer Middleware Framework for Resilient Persistent Programming

Alan Dearle, Graham N.C. Kirby, Stuart J. Norcross & Andrew J. McCarthy
School of Computer Science, University of St Andrews, St Andrews, Fife, Scotland.
{al, graham, stuart, ajm}@cs.st-and.ac.uk

## ABSTRACT
The persistent programming systems of the 1980s offered a programming model that integrated computation and long-term storage. In these systems, reliable applications could be engineered without requiring the programmer to write translation code to manage the transfer of data to and from non-volatile storage. More importantly, it simplified the programmer's conceptual model of an application, and avoided the many coherency problems that result from multiple cached copies of the same information. Although technically innovative, persistent languages were not widely adopted, perhaps due in part to their closed-world model. Each persistent store was located on a single host, and there were no flexible mechanisms for communication or transfer of data between separate stores. Here we re-open the work on persistence and combine it with modern peer-to-peer techniques in order to provide support for orthogonal persistence in resilient and potentially long-running distributed applications. Our vision is of an infrastructure within which an application can be developed and distributed with minimal modification, whereupon the application becomes resilient to certain failure modes. If a node, or the connection to it, fails during execution of the application, the objects are re-instantiated from distributed replicas, without their reference holders being aware of the failure. Furthermore, we believe that this can be achieved within a spectrum of application programmer intervention, ranging from minimal to totally prescriptive, as desired. The same mechanisms encompass an orthogonally persistent programming model. We outline our approach to implementing this vision, and describe current progress.

## 1. INTRODUCTION
During the 1980s, a dichotomy emerged between single address-space programs, increasingly written in an object-oriented style, and distributed programs written using socket abstractions. The syntactic gap between distributed and non-distributed programming then began to close with the advent of middleware systems typified by CORBA [25] and later Java RMI [35]. These systems permitted programmers to program with remote objects in the same manner as local objects. However, many differences remained between such distributed object programs and single address space programs. The differences broadly fall into two categories: those concerning the software engineering process and those concerning differences between local and remote semantics. We discuss each of these in turn.

Industry-standard middleware systems—CORBA, Java RMI, Microsoft COM [20], Microsoft .NET remoting [24] and Web Services [37]—are complex, making the creation of distributed applications difficult and error-prone. Programmers must ensure that application classes supporting remote access correctly adhere to the engineering requirements of the middleware system in use, for example, extending certain base classes, implementing certain interfaces or handling distribution-related error conditions. This affects inheritance relationships between classes and often prevents application classes from being remotely accessed if their super-classes do not meet the necessary requirements. At best, this forces an unnatural or inappropriate encoding of application semantics because super-classes are often required to be accessible remotely for the benefit of their sub-classes. At worst, application classes that extend pre-compiled classes cannot be made accessible remotely at all.

The above systems all require programmers to follow similar steps in order to create remotely accessible classes. Programmers must specify the interfaces between distribution boundaries, and then decide which classes will implement them. Thus classes are hard-coded at the source level to support remote accessibility; programmers must therefore know how the application objects will be distributed at run-time when defining classes—early in the design cycle.

The semantic differences between local and remote programs have been widely discussed in the literature; many regard *A Note on Distributed Computing* [12] as the definitive discussion of these differences. This classifies as *local computing* (local object invocation) those programs that are confined to a single address space, and as *distributed computing* (remote object invocation) those programs that make calls to other address spaces, possibly on remote machines. [12] states that the differences between local and distributed computing lie in four distinct areas: latency, memory access, partial failure and concurrency. [12] argues that since remote invocation is between four and five orders of magnitude longer than a local call, lack of attention to distribution in the design cycle can lead to performance problems that cannot be rectified. Since only local memory can be addressed from within an address space, remote references (pointers) are inherently different from local references. In a single address space program, there is no partial failure; the entire program either fails or runs to completion. By con-trast, in a distributed program, arbitrary components can fail, leading to unpredictable results. Finally, a distributed program exhibits true concurrent behaviour, leading to indeterminacy in the order of invocations.

A rebuttal of these arguments appears in [32], in which Spiegel argues that none of these arguments stand up to close inspection. Spiegel states that the latency argument hinges on the assumption that there must be a static decision on which objects (and classes) are remote and which objects are clustered together. Whilst this is true of early middleware systems, it is not true of some second and third generation middleware systems such as ProActive [6] and RAFDA [9, 16, 28, 38]. Given appropriate middleware support, it is possible to dynamically change the object clustering to avoid latency problems.

High level programming languages such as Java and C# abstract over direct memory access making the remote references issue a non-problem. Similarly with the issue of concurrency, most modern operating systems support multi-threading and newer CPUs support multiple processor cores within what is traditionally thought of as a single CPU. Thus true concurrency exists within a single machine environment. Furthermore, the advent of threading concepts in modern programming languages mean that programmers have to routinely deal with concurrency.

Spiegel argues that there are partial solutions to the problem of partial failures. He argues that in systems with explicit failure detection, the only viable option in the majority of cases is to shut down the program with an error message. For example, with Java RMI, the application programmer is obliged to handle occurrences of *RemoteException* and its subtypes. He argues that while sophisticated retry and replication schemes that could mask the failure are too complicated to be implemented within the application logic, there is a need for infrastructures that can mask failures and shield the application logic. We pick up this mantle here.

Waldo et al. attempt to dispel the vision of unified objects; Spiegel points out flaws in their arguments. In this paper we extend the argument of Spiegel and assert that *it is possible to engineer unified object systems which enable distributed programs to be constructed that are superior to single address space solutions*, in terms of application availability, probability of successful completion, and scalability with respect to storage and compute cycles.

We believe that a distributed system can be engineered to be more reliable than a centralised application. In fact, we assert that it is possible to make use of distribution and replication of active and passive state to provide arbitrary application resilience. Furthermore, we believe that this can be achieved within a spectrum of application programmer intervention, ranging from minimal to totally prescriptive, as desired. In the minimal intervention scenario, the programmer writes an application and the infrastructure makes appropriate choices about active and passive object replication and placement. Conversely, the programmer may wish to exert explicit control over where objects are placed, the manner in which they are replicated, how recovery is performed, etc.

It is important to be clear about the type of application resilience that we address. We are concerned with resilience in the face of failures in the computing infrastructure, including power failures, disk failures, operating system crashes, network outages, etc. We do not address semantic errors: we assume correctness of the core application logic.

In this paper, we explore the building blocks necessary to achieve this application resilience, and outline our approach to implementing such a system by exploiting facilities from the domains of persistent programming, peer-to-peer (P2P) architectures and code transformation.

## 2. VISION

Like the Aspect Oriented Programming community [13], and in contrast to [12], we believe that it is possible to separate persistence, distribution, replication and resiliency concerns completely from the core application logic. These orthogonal aspects may be addressed separately, with the consequence that they may be changed more easily since re-engineering of the application is unnecessary.

Our vision is to support an application development methodology as follows. The application is initially designed, implemented and tested without taking any account of how it will be distributed. The application code is then transformed automatically such that all references between objects become *abstract*. Rather than referring to a specific extant object in the same address space as the reference holder, each reference refers to an abstract object identity.

When accessed, a reference's abstract identity is transparently and automatically resolved—to a local object, to an object in a remote address space, or to an object reconstructed from its flattened object representation, stored on a local or remote disk. A remote reference mechanism allows an object in a remote address space to be used in the same way as a local object. An instantiation mechanism allows an object to be re-instantiated on demand from a flattened representation. This flexibility is used to support several distinct, though related, features:

- Objects are referenced in a location-independent fashion, facilitating dynamic flexibility in distribution topology; an object may be migrated between address spaces without the need to inform reference holders.
- Objects are automatically replicated, with transparent fail-over to a replica when required; reference holders need not be aware of the failure of a referenced object.
- Orthogonal persistence allows objects to outlive a particular run of an application, without the need for the programmer to write any explicit save/restore code.

- Object histories are available, since replicas of previous states may be preserved in the infrastructure.

Executing code may obtain and use an abstract reference to an object without being aware of where on the network that object exists, or indeed whether the object exists in memory or is stored in quiescent form on disk. This is achieved via a discovery service that maps globally-scoped logical names to abstract references. Similarly, code may be passed an abstract reference as a parameter; it is again able to treat that reference in the same way as a local reference. If the referenced object later migrates to another address space, or is flushed to disk and then discarded from memory, this is transparent to the holder of the abstract reference. Information about the most recently resolved target of an abstract reference is cached, so that the resolution cost is not incurred on subsequent accesses, unless the target changes as in the above examples.

Abstract references are resolved using a *key-based routing* service provided by a P2P infrastructure. This provides a scalable decentralised mechanism for locating active objects and flattened representations anywhere in the network.

At intervals determined via a flexible policy framework, the current state of each object is reified and distributed, to form one or more remote replicas.

If the cached remote target object of an abstract reference becomes unavailable due to failure of the remote host or the intervening network, a copy of the object's replicated state is located via the P2P infrastructure, and instantiated to give a new target. This fail-over process is transparent to the reference holder, beyond the additional delay and possible roll-back of the object's state.

The replication infrastructure also supports orthogonal persistence [2] [4]. At the end of a program execution, those objects reachable from a designated root of persistence are replicated on local and/or remote disk. A subsequent execution of the same or another program, whether in the same physical location or elsewhere, can retrieve an object previously made persistent, by obtaining and resolving an abstract reference via the distributed discovery service.

By default, replicas of object states are preserved indefinitely. This means that object version histories are available, allowing the programmer to recover the previous state of an object as recorded at any replication point. The storage cost can be influenced via policy frameworks that control the frequency and granularity of replication, as mentioned above, and the deletion of old replicas (if at all). Although not required, the programmer may also optionally exert control over a number of other policy dimensions, including:

- the initial partitioning of application objects across the network;
- policies for parameter passing semantics;
- recovery.

To summarise, our vision is of an infrastructure within which a conventional application can be deployed and distributed with minimal modification, whereupon the application becomes resilient to certain failure modes. If a node, or the connection to it, fails during execution of the application, the objects are re-instantiated from distributed replicas, without their reference holders being aware of the failure. The same mechanisms encompass an orthogonally persistent programming model. Our aim is to provide an abstraction to the programmer of a global, ubiquitous, reliable, permanent single address-space.

# 3. BACKGROUND

This section describes a number of technologies that we have developed previously, which are exploited in the infrastructure presented in this paper.

## 3.1 Orthogonal Persistence

The essential concept behind persistent programming is that all data values within a programming context are created in an address space that is conceptually permanent and shared among applications [2]. This means that long-term, typed data can be shared among independently compiled units and relieves programmers of writing translation code to manage the transfer of data to and from non-volatile storage (e.g. a file or a database). More importantly, it simplifies the programmer's conceptual model of an application, and it avoids the many coherency problems that result from multiple cached copies of the same information. To quote from [22]:

> *Persistence is used to abstract over the physical properties of data such as where it is kept, how long it is kept and in what form it is kept, thereby simplifying the task of programming. The benefits can be summarised as:*
>
> - *improving programming productivity as a consequence of simpler semantics;*
> - *avoiding ad hoc arrangements for data translation and long term data storage; and*
> - *providing protection mechanisms over the whole computational environment.*
>
> *The persistence abstraction is designed to provide an underlying technology for long-lived, concurrently accessed and potentially large bodies of data and programs.*

In an orthogonally persistent system, any data value can be made persistent, without exception, regardless of its type, how and when it was created, etc. PS-algol [3] was the first language to have orthogonal persistence, while Napier88 [23] was the first language to model persistence within a sophisticated typing regime, including parametric polymorphism, existential data typing, and controlled dynamic typing within a static context.

Although technically innovative, these languages were not widely adopted, perhaps due in part to their closed-world model. Each persistent store was located on a single host, and there were no flexible mechanisms for communication between, or transfer of data between, separate stores.

Since the programmer may treat transient and persistent objects in exactly the same way, any reference encountered during computation may refer to a local in-memory object or to the stored representation of a persistent object on disk. A fundamental implementation requirement for a persistent system is thus support for read barriers. Each reference must be checked before use, and if necessary the referenced persistent object must be faulted from disk and instantiated in main memory. Various 'pointer swizzling' techniques attempt to optimise this process [11].

Conversely, new and modified persistent objects must be written back from main memory to disk. At the minimum, the infrastructure needs to be able to distinguish, on program termination, which objects are new and should now be made persistent, which are already persistent, and of those, which have been modified since being faulted in. Various schemes for identifying those objects that should become persistent have been proposed; in orthogonally persistent systems this is usually achieved by tracing reachability from some root of persistence.

It is also desirable for the system to be able to evict persistent objects from main memory during computation, in order to free space. These must also be written back to disk, if modified.

## 3.2 The RAFDA Run-Time

The RAFDA Run-Time [9, 16, 28, 38] (RRT) is a middleware system that separates distribution concerns completely from the core application logic. Unlike most middleware systems, the RRT permits arbitrary application objects to be dynamically exposed for remote access. This means that changes to distribution boundaries do not require re-engineering of the application, making it easier to change its distribution topology.

Object instances are exposed as Web Services through which remote method invocations may be made. The RRT has the following notable features:

1. The programmer need not decide statically which classes support remote access. Any object instance from any application, including compiled classes and library classes, can be exposed as a Web Service without the need to access or alter application class source code. This is analogous to orthogonal persistence, where any object instance may become persistent, regardless of its type or method of creation.
2. The middleware integrates the notions of Web Services, Grid Services and Distributed Object Models by providing a remote reference scheme synergistic with standard Web Services infrastructures, and extending the pass-by-value semantics provided by Web Services with pass-by-reference semantics. Specific object instances rather than object classes are exposed as Web Services, further integrating the Web Service and Distributed Object Models. This contrasts with systems such as Apache Axis [1] in which classes are deployed as Web Services.
3. Parameter passing mechanisms are flexible and may be controlled dynamically. Parameters and result values can be passed by-reference or by-value and these semantics can be decided on a per-call basis.
4. When objects are passed by-reference to remote address-spaces, they are automatically exposed for remote access. Thus an object *b* that is returned by method *m* of exposed object *a* is automatically exposed before method *m* returns.

The RRT allows application developers to implement application logic without regard for distribution boundaries, and to separately implement code to define the distribution-related aspects. The developer can either abstract over distribution boundaries, or implement distribution-aware code, as appropriate. A flexible policy framework, together with a library of default policies, allows the developer to exert fine-grained control over distribution concerns if required, but to ignore them if not. Although the RRT is written in Java and is designed to support Java applications, it does not rely on any features unique to Java.

An application object may be exposed for remote access in two ways: through an explicit call by the application, or automatically through transmission of a reference to the object. To explicitly expose an object, the application makes a call to the local RRT infrastructure, passing a local reference to the object, an interface type, and optionally a logical name. The interface type specifies the set of methods that will be exposed for remote access. For flexibility, the object itself need not implement the interface type, so long as it is structurally compatible. The signature of the RRT operation used to expose an object is:

```
void expose(Object objectToBeExposed,
   Class interfaceToBeExposed, String name)
```

Each exposed object is dynamically assigned a globally unique identifier (GUID), which provides object identity within the distributed system. The exposed object may be remotely addressed using the pair *(guid, net_addr)*, where *guid* is the object's GUID and *net_addr* is the network address (IP address and port) of the address-space in which the object resides. Where a logical name was provided at the point of exposure, this name may also be used in place of the GUID. The signatures of the RRT operations used to obtain a remote reference to an object exposed in another address-space are:

```
Object getRemote(SocketAddress rrt, GUID g)
Object getRemote(SocketAddress rrt, String name)
```

The result returned from either of these operations is a remote reference, typed as the interface with which the remote object was exposed. This reference can then be used by the application in exactly the same way as a conventional local reference.

Just as orthogonal persistence allows an object to become accessible from another address-space instantiated later in time, the RRT allows an object to become accessible from another address-space elsewhere on the network. Although an RRT remote reference can be used without knowing the actual location of the object referred to, the remote reference does explicitly encode that network location. If that location later becomes unavailable due to host or network failure, the remote reference will become unusable.

The work described in this paper provides mechanisms whereby an object's global identity, that is its GUID, is the only information required to be able to reliably address that object. It extends the features of the RRT such that:

- a remotely accessible object can be located and bound-to from any location, without having to know the object's current location;
- a reference may resolve to a local object, a remote object or to an object reconstructed from a persistent replica, without any difference between these being apparent to the reference holder;
- a remotely accessible object remains available even if the address-space currently containing that object fails or becomes unreachable.

### 3.3 P2P Infrastructures

Our approach centres on the use of Peer-to-Peer (P2P) mechanisms for the reliable addressing of data, stateful services and application objects. P2P routing overlays [27, 29, 34, 39] offer reliable and highly scalable routing mechanisms that map each key in an abstract key-space to a live host in a network. Such an overlay can be used to construct P2P applications that benefit from location-independent addressing of objects, data or services [7, 29]. In a key-based P2P application, each addressable entity is bound to a value in the key-space, and the node to which a particular key maps holds the entity bound to that key, or its location. An addressable entity is thus located by routing to its key in the overlay network.

While an overlay network provides a highly scalable, reliable, self-repairing routing infrastructure, an application built on top of an overlay must ensure that the addressable entities remain available in the face of host and network failure. If the host responsible for a particular range of the key-space fails, the overlay protocol will re-arrange the network's routing data structures to ensure that those keys continue to route to available hosts. It is the responsibility of the application to ensure that the entities bound to those keys are reliably stored and that sufficient information to locate the entities is held at the appropriate nodes in the overlay. This process requires that replicas of the entities or the location information are appropriately placed on the nodes that will take responsibility for the corresponding keys in the event of failure.

[8] defines an API for P2P overlay systems, suitable for the implementation of a range of applications. This defines a routing API for location-independent addressing and an up-call API via which the overlay announces network topology changes to the application layer. This enables applications to move or copy addressable entities and location information to appropriate nodes in response to changes in the mapping of the key-space to live nodes.

The ASA project [15] is developing an autonomically managed storage system based on P2P routing overlay techniques; a number of technologies resulting from this are used in the work described in this paper. These include implementations of multiple overlay protocols and a P2P application infrastructure that supports the construction of applications using these overlays. Our P2P infrastructure supports the construction of key-based P2P applications that are independent of any particular overlay protocol.

## 4. APPROACH

### 4.1 General Principles

Interaction between applications and the P2P infrastructure that supports the middleware is based on the use of keys to identify programming language objects and versions of those objects. A key is associated with an object when the object's identity, or its flattened state, is published on the network. It is useful to differentiate between two syntactically identical types of keys: Globally Unique IDentifiers (GUIDs) and Persistent IDentifiers (PIDs). In our current prototype, both are represented as 160-bit strings—GUIDs are randomly generated, while PIDs are based on content hashing.

GUIDs encompass the notion of identity in a global setting. A GUID serves to identify an object over all time, irrespective of the state of the object. Not all programming language objects need associated GUIDS—only those that take part in global interactions—thus GUIDs are lazily allocated. An extant object associated with a GUID may be located, if one exists, by looking up that GUID in a distributed data structure known as the *Object Directory*, which maps from GUIDs to object references.

A PID is used to identify the state of an object at some particular time, created via a content hash of the serialised state. Over time, as objects are modified, a sequence of (*PID, state)* pairs is generated, and stored in the distributed *Data Store*. The historical sequence of object states is related to object identities by the distributed *Version Directory*. This is an append-only store that maps from object identity (GUID) to a sequence of PIDs associated with that object. Using the PIDs, the state-change history of an object may be discovered (and its state possibly rolled back or forward).

When an object is re-instantiated from a serialised state, its class must be known in order to perform the deserialisation. This information is recorded in the distributed *Code Store*, which maps from GUID to class. The distributed *Policy Store* is a repository for the various policy choices that dictate the behaviour of the middleware system.

The process of resolving an abstract reference involves use of these various distributed data structures:

- extract GUID held in abstract reference and search for it in *Object Directory*
- **if** an extant instance of object is found (either a local reference, or a remote reference to an instance in another address space)
- **then return** reference to extant instance
- **else**
  - search *Version Directory* for an appropriate PID associated with GUID
  - retrieve serialised state associated with that PID from *Data Store*
  - retrieve class associated with GUID from *Code Store*
  - re-instantiate new instance using class and state
  - record new instance in *Object Directory*
  - **return** reference to newly instantiated instance

An abstraction layer is required to shield the application programmer from this complexity. This is discussed in the next section. Note that RRT functionality (as described in Section 3.2) is used to allow the *Object Directory* to return remote references to extant instances, which can be used by the application in exactly the same way as local references.

## 4.2 The Middleware Interface

The middleware presents two APIs to the application, shown in Figure 1 and Figure 2. The first, *NamedObjectDirectory*, represents the simplest interface against which it is possible to program. Using this, the programmer is freed of any responsibility for replication, coherency or recovery. The methods *getObjectByName* and *associateName* are provided for naming and retrieving objects, serving the same purpose as object naming services in CORBA or Java. The difference is that the objects made available using the *associateName* method may be located using the *getObjectByName* method, irrespective of the longevity of the objects, the processes in which they were created, or the context in which either method was called.

The default behaviour of the *commit* method is to replicate the state of the transitive closure of the object associated with a given name *n* times, where *n* is a per-address-space configuration parameter, initially set to 3. This is intended to give reasonable resilience semantics without impacting too greatly on the application programmer.

```
interface NamedObjectDirectory {
   Object getObjectByName(String name);
   void associateName(String name, Object o);
   void commit(String name);
}
```

**Figure 1. The *NamedObjectDirectory* interface.**

In addition to the *NamedObjectDirectory* interface, the more general purpose *PersistenceInfrastructure* interface is also provided. This provides access to the various distributed data structures briefly described in the last section, permitting a variety of application-specific policies to be written to control the replication, coherency and recovery of objects.

The *NameDirectory* methods *getGuidByName* and *associateName* permit the storage and retrieval of associations between logical names and GUIDs.

The version history mapping each GUID to a sequence of PIDs may be accessed using the *VersionDirectory* methods *getLatestVersion* and *versionIterator*, while new versions are published using *publishVersion*.

The *ObjectDirectory* method *getObjects* supplies the application level with a reference to the extant instances associated with the specified GUID. The *getGuid* method returns the GUID associated with an object, allocating a new GUID if necessary, while *getCreationTime* returns the time at which a given GUID was allocated. A new object cannot be accessed remotely until it is made globally available using *publishInstance*, which advertises the object's existence in the *Object Directory*.

```
interface PersistenceInfrastructure {
   NameDirectory getNameDirectory();
   VersionDirectory getVersionDirectory();
   ObjectDirectory getObjectDirectory();
   DataStore getDataStore();
   CodeStore getCodeStore();
   PolicyStore getPolicyStore();
}

interface NameDirectory {
   GUID getGuidByName(String name);
   void associateName(String name, GUID guid);
}

interface VersionDirectory {
   PID getLatestVersion(GUID guid);
   Iterator versionIterator(GUID guid);
   void publishVersion(GUID guid, PID pid);
}

interface ObjectDirectory {
   Object[] getObjects(GUID guid);
   GUID getGuid(Object o);
   Date getCreationTime(GUID guid);
   void publishInstance(GUID guid);
}

interface DataStore {
   Data getObjectData(PID pid);
   void store(PID pid, Data data);
   PID generatePID(Data data);
   Date getCreationTime(PID pid);
}

interface CodeStore {
   Class getClass(GUID guid);
}

interface PolicyStore {
   void setResiliencePolicy(
      Class c, ResiliencePolicy p);
   void setResiliencePolicy(
      Object o, ResiliencePolicy p);
   // ... other policy hooks omitted
}

interface ResiliencePolicy {
   Data reify(GUID guid);
   Object instantiate(GUID guid);
   PID makeResilient(GUID guid);
   // ... other resilience policy hooks omitted
```

```
}
```

**Figure 2. *PersistenceInfrastructure* and related interfaces.**

The data associated with a PID is retrieved using the *DataStore* method *getObjectData*. This returns an instance of *Data*, which is an abstraction over the unstructured data (bytes) holding the object's serialised state. The *store* method initiates the replication of the supplied data, keyed by a given PID, while *generatePID* creates a PID for the given data. The method *getCreationTime* returns the time at which a given PID was generated.

The *CodeStore* method *getClass* retrieves the class associated with a given GUID. Finally, the *PolicyStore* methods provide hooks for associating application-specific policy with particular classes or objects. The next section describes how this may be used to control resilience.

## 4.3 Controlling Policy

As mentioned in Section 2, one of our goals in developing the infrastructure described in this paper is to allow applications to be deployed on the middleware with minimal change, while at the same time making it possible for the programmer to exert fine-grained control over the operation of the middleware if required. Extending our RRT work, our approach is to identify the dimensions suitable for the application of user-level policy, and then to provide a framework that allows the programmer to specify particular policies, together with default policies, which are designed to be satisfactory for simple applications. To retain flexibility, each policy choice may be specified independently of the application code, and may be changed dynamically. Depending on the policy aspect, policies may be associated with classes, methods, individual method parameters, or with particular objects. We now give an overview of the policy aspects that may be controlled, with examples of how particular choices may be specified.

### *4.3.1 Resilience Policy*

Resilience is achieved through automatic object replication and recovery. A number of dimensions of the object replication may be configured, including:

- application-level consistency requirements
- whether replica propagation is performed synchronously with respect to the application code
- the number of replicas to be distributed, and constraints on their placement, including geographical and whether in volatile or non-volatile storage
- the mechanisms used to transmit the replicas, including serialising the entire current state, encoding a delta relative to a previous state, sending a code fragment to carry out the update, etc
- the format in which the replicas are stored, including a direct serialised form, erasure encoding etc
- whether replicas of old versions should ever be deleted, and if so, when

These dimensions can be controlled by associating specified policy objects with application classes or instances using the *ResiliencePolicy* interface. We first consider the default resilience policies, which take no account of any application-specific consistency requirements.

#### *4.3.1.1 Replication Policy*

Various replication policies may be composed from the dimensions listed above, including extremes such as disabling replication altogether—giving minimal overhead with minimal resilience—and synchronous replication on every field update—giving maximal resilience but with higher overhead.

The default replication policy is for every object within the transitive closure of a named object to be replicated whenever the commit method from the *NamedObjectDirectory* interface (Figure 1) is invoked on that name. The implementation of *commit* is fixed; it simply invokes the *makeResilient* method from the *ResiliencePolicy* interface on the GUID corresponding to the given name. This call is controlled by the resilience policy currently in effect in that context. The default implementation of *makeResilient* is shown in Figure 3:

```
PID makeResilient(GUID guid) {
   Object obj = objectDirectory.getObject(guid);
   for each Object o in closure of obj {
      Data data = reify(o.getGuid());
      PID pid = dataStore.generatePID(data);
      dataStore.store(pid, data);
      versionDirectory.publishVersion(guid, pid);
   }
   return PID generated for initial object obj;
}
```

**Figure 3. Default *makeResilient* implementation.**

The first action of the method is to retrieve the object corresponding to the given GUID from the *Object Directory*. For brevity, details of access to this and the other distributed data structures are omitted; the expression:

```
objectDirectory.getObject()
```

is used as a short-hand for:

```
Infrastructure.getPersistenceInfrastructure().
   getObjectDirectory().getObject()
```

The *makeResilient* method then traverses the transitive closure of the given object. Every object encountered is converted to a flattened representation of its state with a call to *reify*. Again, this call is controlled by the resilience policy currently in effect. A PID is generated by hashing the flattened state with a call to *generatePID*. The appropriate number of replicas is propagated to other nodes in the network via *store*. Finally, the new state of the object is published via *publishVersion*.

The action of replicating an object is equivalent to flushing or checkpointing a persistent object to a persistent store; in both cases the object can then be discarded from volatile memory, since it can be re-instantiated from non-volatile state if needed again. Thus the specification of replication policy can also be thought of as controlling the system's orthogonal persistence functionality.

Where this replication policy is unsuitable for an application, the programmer may exert finer control by specifying a customised implementation of *ResiliencePolicy* to be associated with an application class or with a specific object. Such customised versions may use any of the functionality provided via the *PersistenceInfrastructure* interface. The *makeResilient* method in such a policy might, for example, omit the traversal of a particular part of the object closure, if it is known that that part of the graph should be treated as volatile and not made persistent. Another example, concerning application-level consistency, is discussed in the next section.

It is, of course, possible that a node or network failure may occur during an execution of *makeResilient*. To reduce the probability of inconsistencies arising, the following properties are guaranteed by the infrastructure:

- The storage of an individual replica is atomic: it will either be stored completely or not at all. Thus the effect of *store* is to store some number (possibly zero) of complete replicas on the network.
- The overall effect of *commit* is atomic: it will either succeed in replicating the complete closure or will have no externally visible effect. Thus the closure retrieved by calling *getObjectByName* from another node on the network always corresponds to an explicitly committed state.

*4.3.1.2 Re-instantiation Policy*

It may be necessary to re-instantiate an object from flattened stored state in two situations: when the failure of a live object is detected, and when access is made to a persistent object not currently held in memory. Whenever an object needs to be re-instantiated, the infrastructure invokes the *instantiate* method from the *ResiliencePolicy* interface on the corresponding GUID. This call is controlled by the resilience policy currently in effect in that context. The default implementation of *instantiate* is shown in Figure 4:

```
Object instantiate(GUID guid){
   PID version =
      versionDirectory.getLatestVersion(guid);
   Data data = dataStore.getObjectData(version);
   Class c = codeStore.getClass(guid);
   Object o = instantiateObject(data, c);
   objectDirectory.publishInstance(guid);
   return o;
}
```

**Figure 4. Default *instantiate* implementation.**

The method first retrieves the most recent PID from the *Version Directory*, and a serialised state replica corresponding to that PID from the *Data Store*. The class of the object is retrieved from the *Code Store*, and used to re-instantiate the object from the replica, in the local address-space. Finally, existence of the new instance is advertised in the *Object Directory*.

The programmer may register customised recovery code by associating an alternative implementation of *instantiate* with an application class or with a specific object. Aspects that could be customised include the location of the re-instantiated object—this could be in a different address space from that initiating the recovery—and policy controlling the number of instantiations of the object that may co-exist.

The operations *makeResilient*, *reify* and *instantiate* are interdependent, hence their grouping in the *ResiliencePolicy* interface. Since they operate on unstructured data (*Data*), the manner in which state is represented is divorced from the infrastructure supporting that state. The infrastructure has no knowledge of how persistent data is stored—it could be as XML, Java serialised format, or some optimised format taking advantage of application domain knowledge. The application programmer can control this by defining customised implementations of *ResiliencePolicy*.

*4.3.1.3 Customised Resilience Policies*

Several example uses of customised resilience policies have already been mentioned. Another arises with respect to application-level consistency, for example in situations traditionally addressed by ACID transactions. The atomicity of the *commit* operation is not sufficient to ensure atomicity of concurrently executing programs.

For example, consider the archetypal banking example, in which two concurrent or interleaved operations each make a transfer between two accounts. Figure 5 shows a simple data structure in which a named root “bank root” refers to an object representing a bank. This, in turn, refers to objects *A*, *B*, *C* and *D*, instances of class *Account* representing individual accounts.

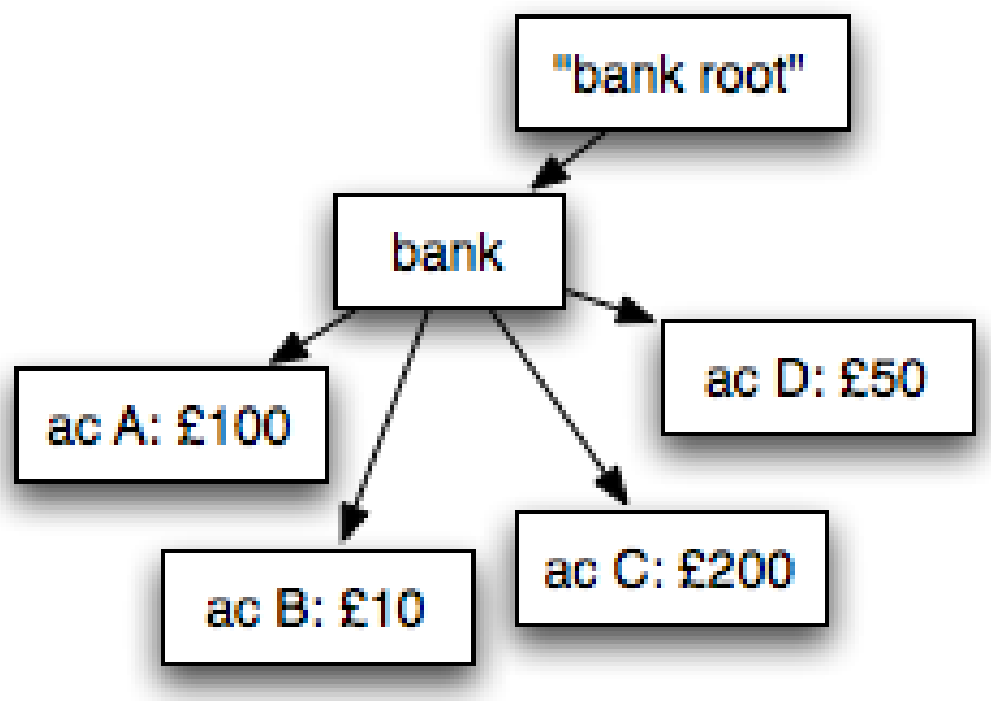

**Figure 5. Example data structure.**

Assume that threads *1* and *2* are executed concurrently to perform the two transfers:

```
thread 1:
  get bank object via getObjectByName("bank root")
  subtract £10 from account A
  add £10 to account B
  commit("bank root")

thread 2:
  get bank object via getObjectByName("bank root")
  subtract £10 from account C
  add £10 to account D
  commit("bank root")
```

A problem arises if the threads are interleaved such that the subtraction in thread *2* has been performed by the time that thread *1* completes, and thread *2* is then killed (or the node on which it is running crashes) before completing. In this situation the *commit* operation performed by thread *1* will make resilient the bank data structure as it was at the time of completion of thread *1*. The newly resilient data structure includes the updated state of account *C*, but not that of account *D*, since thread *2* did not complete its *commit* operation. The result is an inconsistent data structure, as shown in Figure 6, in which the intended invariant (a transfer between accounts should not alter the sum of the balances) has not been preserved:

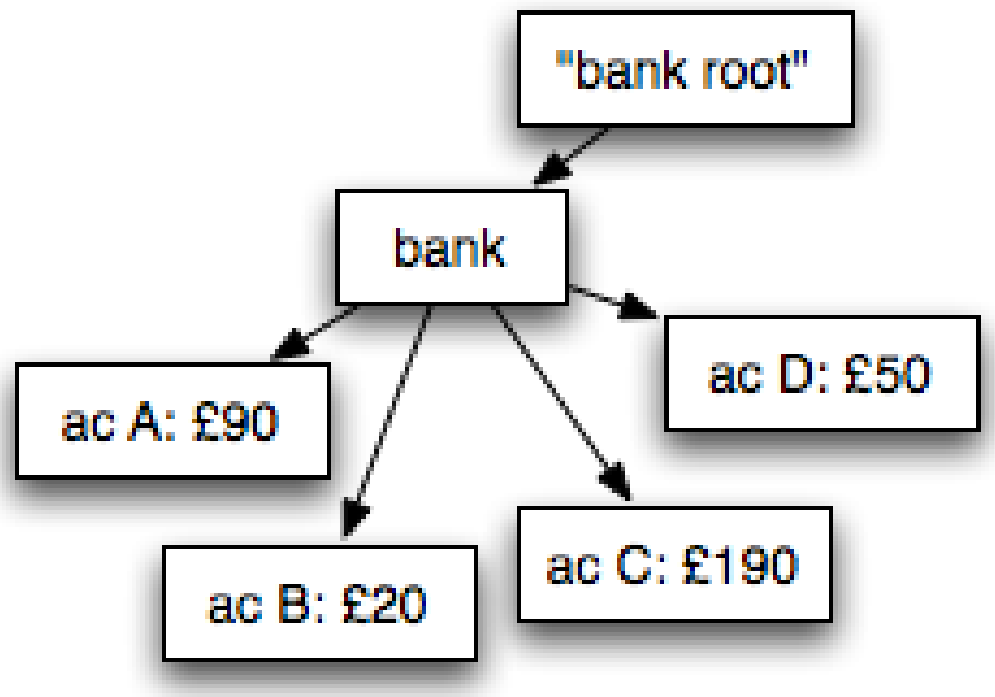

**Figure 6. Inconsistent state.**

The problem in this example is due to the atomicity of update to the persistent state of the entire object closure, rather than the desired atomicity of the individual transfer operations—the update operations of thread *2* are not performed atomically, since the first update is made persistent but not the second. A similar problem arises in situations where multiple top-level roots share common sub-structure: a commit performed on one has a (potentially unexpected) effect on the others.

Such problems are well known; they have been addressed in database and programming language systems by ACID transactions. There is also a large body of work on more flexible non-ACID approaches to consistency control. No single set of semantics is suitable for all applications; hence, we avoid building-in even ACID properties to our middleware. Instead, our approach is to allow the application programmer to implement appropriate policies, when required, using the functionality provided through the *PersistenceInfrastructure* interface[1].

We now examine how the example problem can be addressed via customised policies, by implementing transactional semantics at the application level. A customised replication policy for the bank data structure can be established by associating an alternative implementation of *makeResilient* with the root object, and disabling the default policy for the *Account* class. A simple optimistic scheme could be implemented as follows:

- associate a transaction identifier with each account update operation
- maintain a data structure recording the identifiers of the transactions updating each account object
- in the customised version of *makeResilient*, traverse the object closure to determine which account objects have been modified by this transaction
- if none of these account objects has also been modified by another transaction, propagate replicas for all the modified objects and publish the new versions
- otherwise abort the transaction by returning immediately

[1] This approach is similar to that proposed in [33].

### *4.3.2 Other Policy Aspects*

Beyond resilience, several other aspects of the middleware can be controlled by application-specific policy.

The initial partitioning of an application is determined by object placement policy. This specifies the address spaces in which new objects should be instantiated (in the same way that recovery policies may specify locations for re-instantiation). The default is for new objects to be instantiated in the same address-space as the caller.

Fine control may be exerted over parameter passing semantics [38]. Pass-by-value, pass-by-reference and hybrid schemes can be specified dynamically for classes, methods and individual parameters. The default policy is pass-by-reference.

# 5. IMPLEMENTATION

In this section, we discuss implementation related issues, including the use of P2P abstractions, and the implementation of the six stores of which the system is logically comprised (the version directory, the data store, the object directory, the name directory, the policy store and the code store).

## 5.1 Implementation Overview

Our approach centres on the use of P2P mechanisms for reliably addressing extant application objects, their replicas and flattened object state. The use of P2P is attractive since they offer a highly scalable, reliable, self-repairing routing infrastructure. Here we describe the P2P based addressing and storage mechanisms that support the implementation of our middleware system.

As described above, the storage model logically consists of six storage categories corresponding to the *Version Directory*, the *Data Store*, the *Object Directory*, the *Name Director*, the *Policy Store* and the *Code Store*. Each of these stores is implemented as a decentralised service hosted on a P2P overlay network. A set of service objects is instantiated on each of the nodes of the overlay with each node providing service objects for all six storage categories. Each node in the infrastructure is responsible for some range of keys $[k_x,k_y]$. Each service object hosted on a node is responsible for the storage of objects that map to the key range of the hosting node. Thus if node $N_1$ is responsible for some range of keys $[k_x,k_y]$ then the *Version Directory* service object on node $N_1$ holds the version history for all GUIDs in the range $[k_x,k_y]$, the *Data Store* service object on node $N_1$ records the serialised state of objects whose keys are in the range $[k_x,k_y]$ and so on.

In order to address each of these storage service objects it is necessary to be able to differentiate between them. Each storage category is associated with a distinct *Application Identifier* (AID). The AID is used by a decentralized object location method called *dol*, provided by every point of presence that offers access to the P2P infrastructure:

```
public Object dol(Key k, AID serviceID);
```

The *dol* method permits the caller to obtain a reference to the object that provides a particular service on the P2P overlay and is responsible for the range of the key space in which key *k* lies. Thus, a *dol* call for a key *k* with the *Version Directory* AID will return a reference to the object implementing the *Version Directory* service on the node in the P2P overlay responsible for the key range containing *k*. The effect of this key-based approach, assuming a random distribution of allocated keys, is to spread the load among the nodes of the overlay and consequently the instances of the storage components deployed on the overlay. To carry out any operation on one of the components, the client executes a *dol* call to obtain a reference to the appropriate object in the network and then calls the required method on that object.

## 5.2 Generic Reliable Storage

All the storage categories have broadly similar storage requirements and therefore make use of generic common storage service objects located on each node. Providing generic common storage has the benefit of being able to manage the replication and resiliency of data and metadata in a single place. The generic storage interface is shown in Figure 7. This interface is locally available to all the storage service objects located on a P2P node. It is also exposed to the network to allow for the storage, update and retrieval of replica data and metadata.

```
interface GenericStore {
   void put(Key k, Data data);
   Data get(Key k);
   Data update(Key k, Data data);
   void append(Key k, Data data);
   Data remove(Key k);
   Iterator getAll();
}
```

**Figure 7. The *GenericStore* interface.**

All of the common P2P overlay abstractions [27, 29, 34, 39] provide resilient routing in the face of node failure and topology change. However, changes to the set of nodes hosting the storage service objects impact the data storage provision. For example, a new node may become the primary node for data already stored on another node. Consequently, some existing data may have to be copied onto the new node serving as its primary node. Similarly, data may have to be replicated further, as nodes holding replicas of data leave the P2P network. Thus, the data storage layer needs to have knowledge of changes in the ring topology. To accommodate this need, the P2P layer provides an up-call mechanism, which informs the generic storage system and other high-level components of changes in the P2P topology. On reception of the up-call, the store reconfigures the data that it holds in order to effect repair with respect to the number of extant copies of the data and their locations.

### *5.2.1 Storage Policy*

We have explored a number of implementation strategies for the data storage service. The first strategy is to colocate the storage of data associated with some key with the data storage service object responsible for the key-range in which that key lies. Using this strategy, the set of nodes on which data is stored is determined by the topology of the P2P overlay. We illustrate this using the *DataStore* interface shown in Figure 2. With this interface, no flexibility exists with regard to the placement of data within or outwith the P2P overlay. The data is always stored on the node offering this service interface and on the other nodes chosen by this node on which to store replicas.

A second approach is to allow data to be stored on arbitrary hosts and to record the locations of the data in the storage objects deployed on the overlay network. Using this strategy, the storage service located on a P2P node records the network locations of all instances of data with keys that lie in that node's key-range. To permit this flexibility, additional methods are required that permit the service objects to be informed of the location of the data and for clients to later retrieve it. These additional methods are illustrated in Figure 8, in which the *GenericStore* interface is reused to describe storage services provided by arbitrary hosts. Such storage hosts need not be part of the P2P overlay itself.

```
GenericStore[] getStore(PID pid);
void recordDataLocations(PID pid,
   GenericStore[] repositories);
```

**Figure 8. Additional *DataStore* interface methods**

Using these methods, client code can decide where to store data and how many copies to make. Thus, using this approach, the job of the storage components in the overlay is to resiliently record the locations of the copies; this is effectively a discovery service. The benefit of this approach is that the storage policies are separated from the policy for making the discovery services resilient using the P2P overlay. This additional flexibility comes at a cost—the overlay cannot guarantee that the data is resilient, since this responsibility is assumed by the client middleware.

# 6. RELATED WORK

## 6.1 Replication

The replication of processes and data is widely used to increase availability, performance and fault-tolerance. In distributed file systems such as Coda [30] replication is used to increase the availability of data. Clients of the Coda file system transparently communicate with a set of replicated servers, which provide a level of fault tolerance and may perform local caching to facilitate disconnected operation. In the event of failures, Coda does not provide any guarantees of consistency other than ensuring any inconsistent replicas will be identified after the failure is resolved and made available for manual resolution.

More recently, OceanStore [17] aims to provide "Global-Scale" persistent storage designed to operate over an untrusted infrastructure where servers are unreliable and may not be available; servers are not trusted and may leak information to unauthorised parties. OceanStore is built upon the Tapestry [39] decentralized object location and routing system (DOLR). In OceanStore, objects are globally identifiable via their GUID, and consist of a number of distinct versions, each identified by a version GUID (VGUID). OceanStore employs a two-fold approach to consistency and replication management. Firstly—related to the approach described here—changes to an object's state result in a new read-only version of the state being created and assigned a VGUID (analogous to our PID); this state is then replicated. Secondly, OceanStore utilises primary-copy replication, in which each object has a single primary replica that manages all updates to the object, propagating changes by publishing a signed certificate mapping the object's GUID to the latest VGUID.

Instead of replicating only data to increase availability and fault tolerance, an alternative approach is to replicate an application's active components. One example of this approach is JGroup [21], which allows clients to communicate with a group of active replica objects as though they were communicating with a single conventional Java RMI [35] server object. Should the failure of one of the replicas in the group occur, the client will remain unaware and the infrastructure will allow the client to automatically communicate with another suitable replica. JGroup's Autonomous Replication Management (ARM) [19] allows a group of replicas to be associated with a customisable level of redundancy. During normal execution, ARM monitors the group of replicas and ensures that a specified level of redundancy is maintained. Should the current level of redundancy fall, ARM will take appropriate steps to restore the redundancy level, for example by instantiating new replicas. A framework for the creation of application specific state merging protocols is provided, which can be used to re-establish consistency within an object group after a network partition occurs.

## 6.2 Recovery Oriented Computing

The central thesis of Recovery Oriented Computing (ROC) [5] is that any conventional application and particularly any distributed application will experience a failure during its normal execution which will impede or destroy its ability to continue fulfilling its design purpose. Such failures may be due to internal causes such as software errors and resource exhaustion, or may be due to external factors including network/power outages, hardware failures, security breaches or human errors. ROC recognises that such failures are inevitable and in order to build dependable and highly available systems, such software must be built to quickly recover from failures.

The key concept behind ROC is the "Three R's": Rewind, Repair and Replay. In essence, after a catastrophic failure has occurred, the stricken system can be 're-wound' to a state which corresponds to the state of the system prior to the failure. The failure can then be pre-empted and the execution of the system is "replayed" or restarted, and provided with all previous inputs to the system as before. However the system no longer suffers from the catastrophic failure and can therefore continue to execute normally.

## 6.3 Middleware

The difficulties inherent in creating and configuring distributed applications using common middleware systems were described in the introduction. These difficulties are addressed by several second-generation middleware systems, which allow programmers to employ code transformation techniques to generate distribution-related code automatically. J-Orchestra [36] and Pangaea [32] transform non-distributed applications into distributed versions based on programmer input. They perform static code analysis and employ tools to help programmers choose suitable partitions. Distributed versions of applications are automatically generated from the local versions and so the re-engineering process is simplified, making a trial and error approach to creating applications more feasible.

ProActive [6] and JavaSymphony [10] allow objects to be exposed to remote access dynamically. However, both subtly alter application threading semantics and force programmers to ensure referential integrity manually through their use of active objects [18]. This requires programmers to consider both application distribution and the middleware system's threading model at class creation time in order to ensure that thread safety is retained after objects are exposed to remote access or migrated to other address-spaces.

No current middleware systems, however, support location-independent addressing of arbitrary application objects, with automatic fail-over to remote replicas, as described in this paper.

## 6.4 Persistence

Many current systems aim to provide persistent object functionality. Here we examine three of them: the CORBA Persistent Object Service, Enterprise Java Beans and Aspect-Oriented Computing.

The aim of the CORBA Persistent Object Service (POS) is *to provide common interfaces to the mechanisms used for retrieving and managing the persistent state of (CORBA) objects* [26]. The CORBA POS is composed of several (independent) abstractions that combine to provide a service:

- a *Datastore* provides a particular mechanism for maintaining an object's persistent state;
- a *Persistent Identifier* identifies the location of an object's persistent data in a *Datastore*;
- a *Persistent Object* is an object that supports an interface allowing a client to control the persistence of its state;
- the *Persistent Object Manager* redirects the abstract persistence requests from a POS client to a particular mechanism used to control an object's persistence; and
- the *Persistent Data Service* provides an interface that applies a protocol to a persistent object in order to store its state in a particular *Datastore*.

While these abstractions relate to the ones we propose here, the POS has a number of associated problems. Prin-cipally, there is no failure model, resulting in applications that cannot reason about failure, and there is a lack of control mechanisms that allow for the persistence and recovery of compound objects.

Enterprise Java Beans (EJB) support two styles of persistence: *Container Managed Persistence* and *Bean Managed Persistence*. Using *Container Managed Persistence*, an entity bean relies on its container to manage the transfer of data between the entity bean instance's variables and the underlying resource manager (database). An entity bean with container managed persistence must not code explicit data access: all data access must be deferred to the container. Kienzle and Guerraoui [14] point out that using *Container Managed Persistence* the container does not have any knowledge of the semantics of the methods of a bean, and therefore must make a "blind" choice when implementing concurrency control (and persistence). This is highly inefficient. By contrast, an entity bean utilising *Bean Managed Persistence* is responsible for managing its own state stored in an underlying database. Using *Bean Managed Persistence*, the entity bean provider typically writes database access calls using JDBC or SQL directly in the entity bean component. The commonality with this paper is that the EJB standard specifies a protocol consisting of a number of states that define the lifecycle of beans, giving the bean provider a clear understanding of their responsibilities with respect to object lifetime.

The approach described in this paper has much in common with the Aspect-Oriented Programming (AOP) approach. Persistence is one of the much cited aspects that can be addressed using AOP techniques. Kienzle and Guerraoui [14] examine the relationship between AOP and transactions, concurrency and failures. Their OPTIMA framework supports optimistic and pessimistic concurrency and a variety of different recovery strategies. They observe that 'aspectising' transactions is doomed to failure, because of the incompatibility of the linearisability of method invocations provided by shared objects and transaction serialisability. While separating the transactional interfaces from the rest of the program can be achieved using aspect-oriented programming techniques, such separation is artificial since the transactional aspect is actually part of the semantics of the object to which it applies. These observations reinforce our approach of appropriate levels of programmer intervention in persistence, recovery (and transactions).

[31] specifically addresses persistence and distribution aspects. The distribution aspects implement basic remote access to (client-server) services using Java RMI and the persistence aspects implement basic (non-replicated) persistence functionality using relational databases. They state that the (aspect oriented) patterns for persistence and distribution may be encoded in code generation tools and automatically generated for different applications. This is synergistic with the approach taken here.

# 7. IMPLEMENTATION STATUS

The implementation of the work described in this paper is ongoing. The ASA project has implemented a number of P2P routing architectures including CAN, Pastry and Chord with a common API. The distributed directories described here have been implemented against this common API. The ASA project also supports the generic persistent storage architecture described in this paper. The RAFDA system is fully implemented and supports both the P2P routing architectures and the directories implemented above it. The RAFDA system also supports policies controlling various aspects of distribution. We are currently integrating these disparate technologies into the system described here.

# 8. CONCLUSIONS

Our aim is to provide an abstraction to the programmer of a global, ubiquitous, reliable, permanent single address-space. The motivation for this arises from our experience with flexible middleware, P2P systems and persistent programming systems.

The architecture combines aspects of the RAFDA middleware and persistent programming systems. In the former, the programmer can treat references to local and remote objects in the same way, while in the latter, the programmer can treat references to objects in memory and their replicated flattened form stored on resilient storage in the same way. In the work described here, all three kinds of reference are combined into a single unified addressing model.

The persistent systems of the 1980s supported *orthogonal persistence* meaning that persistence was orthogonal to other aspects. Although technically innovative, persistent languages were not widely adopted, perhaps due in part to their closed-world model. Each persistent store was located on a single host, and associated with fixed management policies. By integrating persistence with reliable, replicated P2P storage, data can become truly ubiquitous and independent of any node. Furthermore, by exposing suitable interfaces to the P2P infrastructure, application specific resilience, recovery and transaction policies can be implemented if desired. Thus the system permits a spectrum of application programmer intervention with respect to persistence, distribution and replication, ranging from none as is the case in orthogonally persistent systems to totally prescriptive, which may be desired in highly tuned commercial environments.

We have sketched an architecture that provides an abstraction to the programmer of a global, ubiquitous, reliable, permanent single address-space. This is superior to a non-distributed solution in terms of application availability, probability of successful completion, and scalability with respect to storage and compute cycles. We have demonstrated that it is possible to achieve a useful approximation to this ideal through data replication and self-organising P2P overlays.

## 9. ACKNOWLEDGEMENTS

This work was supported by EPSRC grants GR/R51872 and GR/S44501/01 and by Nuffield grant URB/01597/G.